\documentclass[preprint,12pt,sort&compress]{elsarticle}
\usepackage{lineno,hyperref}
\usepackage{geometry}                		
\usepackage{graphicx}
\usepackage{epstopdf}		
\usepackage{subfigure}
\usepackage{amssymb}
\usepackage{multirow}
\usepackage{caption}
\makeatletter
\def\ps@pprintTitle{%
   \let\@oddhead\@empty
   \let\@evenhead\@empty
   \let\@oddfoot\@empty
   \let\@evenfoot\@oddfoot
}
\newbox\absbox
\renewenvironment{abstract}{\global\setbox\absbox=\vbox\bgroup
  \hsize=\textwidth%
  \noindent\unskip\textbf{ABSTRACT}
 \par\medskip\noindent\unskip\ignorespaces}
 {\egroup}

\begin{document}
\begin{frontmatter}

\title{Magnetic and structural properties of Co$_2$MnSi based Heusler compound}
\author{S. J. Ahmed$^\dag$, C. Boyer$^\sharp$, M. Niewczas$^\dag$$^\ddagger$$^{*}$}
\address{$^\dag$ Materials Science and Engineering, McMaster University, Hamilton, Ontario, Canada\\$^\sharp$Canadian Neutron Beam Centre, Canadian Nuclear Laboratories, Chalk River, Ontario, Canada\\$^\ddagger$Brockhouse Institute of Materials Research, McMaster University, Hamilton, Ontario.\\
$^{*}$E-mail: niewczas@mcmaster.ca}

\date{\today}

\begin{abstract}
The influence of antisite disorder occupancies on the magnetic properties of the half-metallic Co$_2$MnSi compound was studied by experimental techniques and first-principles calculations. The neutron diffraction studies show almost equal amount of Mn and Co disorders of 6.5\% and 7.6\%, which was found to be in good agreement with density functional theory (DFT) calculations of the stable Co$_2$MnSi system with the corresponding disorders. DFT studies reveal that antiferromagnetic interactions introduced by Mn disorder lead to a reduction of the net magnetic moment. The results are discussed in conjunction with neutron diffraction and magnetization measurements. Magnetotransport measurements revealed a positive magnetoresistance for bulk Co$_2$MnSi, which decreases as temperature increases. A Curie temperature of $\sim$1014 K was determined for the compound by high-temperature electrical resistivity and dilatometry measurements.
\end{abstract}

\begin{keyword}
Heusler compounds; Half-metallicity; Antisite disorder; Neutron diffraction; Density Functional Theory; Magnetic properties; Magnetoresistance.
\end{keyword}

\end{frontmatter}

\section{Introduction}

Half-metallicity has been a topic of tremendous scientific attention for the last three decades since the theoretical prediction of a 100 \% spin polarization in the half-Heusler NiMnSb~\cite{DeGroot1983} and later, on the full-Heusler Co$_2$MnSn~\cite{Kuebler1983}. Recent progress on the phenomenon has been extended to systems like Weyl semimetal~\cite{Xu2018,Liu2018,Wang2018a}, Dirac half-metals~\cite{Sun2018,Ma2018,Wang2018}, d$^{0}$-d half-Heusler~\cite{Davatolhagh2018,Dehghan2019}, and also, perovskites~\cite{Kim2018,Hsu2018,Zu2018}. Nevertheless, low Curie temperatures of these systems has kept their room temperature operation still out of reach. In contrast, the transition metal based full-Heusler compounds offer half-metallicity accompanied by high Curie temperatures, scalability of electronic properties, and negligible spin-orbit coupling~\cite{Ishida1995,Inomata2003,Kaemmerer2004,Yakushiji2006,Furubayashi2008,Kodama2009,Nikolaev2009,Graf2011,Pradines2017,Wang2018b,Wang2018c,Siakeng2018}. A 100\% spin polarization, however, has not been achieved experimentally. The discrepancy between the theoretical predictions and experimental results is mainly attributed to the presence of structural disorder and the finite temperature effects\cite{MacDonald1998,Dowben2003,Picozzi2004,Katsnelson2008,Chioncel2008,Pradines2017}. The transition metal elements in these compounds tend to swap sites with each other what introduces new states in the minority spin gap destroying half-metallic properties. The spin disorder that occurs at finite temperatures is another cause of the destruction of half-metallic properties.\\

Within the full-Heusler class, one of the most highly considered candidates for half-metallic ferromagnetism, the Co-Mn-Si based compound Co$_2$MnSi exhibits a large theoretical spin band gap of $\sim$0.4 eV~\cite{Fujii1990} and a high Curie temperature of $\sim$985 K~\cite{Webster1971,Brown2000} that are desired attributes of materials for spintronics. The theoretical and experimental work conducted on Co$_2$MnSi in the last two decades has focused on the analysis of structural and magnetic properties and their relation to spin polarization~\cite{Ishida1995,Kaemmerer2004,Cheng2001,Ravel2002,Raphael2002,Ritchie2003,Sakuraba2005,Singh2004,Picozzi2004,Wang2005,Akiho2013,Jourdan2014,Pradines2017,Rath2018}. The highest reported value of spin polarization, $\sim$93\%, was measured by ultraviolet-photoemission spectroscopy for bulk Co$_2$MnSi at room temperature~\cite{Jourdan2014}. For thin films, spin polarization up to 89\% has been observed at low temperatures where the effect of spin's disorder is minimized~\cite{Sakuraba2005}. Recently, \citet{Moges2016} studied the relationship of spin disorder with the temperature in off-stoichiometric Co$_2$MnSi$_{0.84}$ thin films. The authors reported for this compound a transverse magnetoresistance ratio (TMR) of 1400\% at 4.2 K, compared to 300\% at room temperature.\\

Despite numerous investigations, some aspects of structure-property relationship in Co$_2$MnSi compound need to be better understood. One questions concerns the structural disorder between Co and Mn sites, which influences the properties of the compound and determines the half-metallic behaviour. The site occupancy parameters for Co$_2$MnSi was studied by \citet{Ravel2002} using the neutron diffraction and X-ray absorption fine structure (EXAFS) technique. The results of these authors pointed towards a similar disorder affinity of $\sim$8-14\% for both Co and Mn sites. On the contrary, a theoretical study by \citet{Picozzi2004} reported a difference of disorder affinity more than twice with a findings that Co disorder occurs less likely. Although, the disagreement was suggested to be related to the computational parameters, there has not been any further attempts to address this problem.\\

There are also inconsistencies in the reported magnetic moment of Co$_2$MnSi compound. The neutron diffraction measurements report it in the range between 5.16-5.62 $\mu_{B}/f.u.$ at room temperature~\cite{Brown2000,Ravel2002} and 5.07 $\mu_{B}/f.u.$ at 4 K~\cite{Webster1971}. Ravel and co-workers~\cite{Ravel2002} has recommended that the refinement of magnetic moment requires further investigation. A recent theoretical study by \citet{Pradines2017} and the earlier work by \citet{Picozzi2004} both suggested a reduction of the total magnetic moment in equilibrium conditions at 0 K, due to an antiferromagnetic interactions induced by Mn antisite disorder. These predictions have never been verified experimentally.\\

Another question concerns the transport properties of Co$_2$MnSi under magnetic field. The longitudinal magnetotransport studies of the bulk Co$_2$MnSi by~\citet{Ritchie2003} reported that the system shows no magnetoresistance in the field of up to 5.5 Tesla. This finding however contradicts with the small positive magnetoresistance that is usually exhibited by the other half-metallic systems Co$_2$FeSi~\cite{Bombor2013}, Fe$_{0.8}$Co$_{0.2}$Si~\cite{Manyala2000} and CrO$_{2}$~\cite{Watts2000,Coey2002}.\\

In the present work, we investigate the effect of disorder on the magnetic moment in Co$_2$MnSi half-metallic Heusler alloy. The neutron diffraction studies of the compound were conducted at 298 K, 100 K, and 4 K to retrieve the structural and magnetic parameters. First-principles calculations with defects were performed that validate the experimental observations and explained the correlation of the disorder occupancies and the magnetic moment. Temperature and field dependent transport properties were investigated to elucidate the role of charge transport in the magnetism of Co$_2$MnSi.

\section{Experimental and computational details}\label{sec:Experimental details}

Co$_2$MnSi alloy was prepared by arc-melting of the pure elements (Co (99.95\%), Mn(99.98\%) and Si(99.999\%) ) under argon atmosphere. To improve homogeneity, the ingot was remelted three times. The excess Mn was added to compensate for the evaporation loss of the element. The compound was then sealed in an evacuated silica tube, annealed at 800 $^{\circ}$C for two weeks and subsequently quenched in mixture of ice and water to further improve the crystallinity. No mass loss was observed due to annealing. Phase purity was confirmed by X-ray powder diffraction using \emph{PANalytical X'Pert Pro diffractometer} with Co K$\alpha_{1}$ radiation.  The refining of the diffraction data was performed using the full-profile Rietveld refinement implemented in the FullProf program \cite{Rodriguez-Carvajal1990,Rodriguez-Carvajal2001,Rodriguez-Carvajal2011}. The magnetic configurations was generated with representation analysis program SARAh \cite{Wills2000}.\\

Single crystals of Co$_2$MnSi alloy were grown using the RF heating Czochralski crystal growth method in an argon atmosphere. The pure elements were melted in an alumina crucible and the crystal was pulled using a tungsten wire seed with a constant pulling rate of 0.5 mm/min and 30 rpm rotation. The Laue diffraction revealed that the crystal has grown along $<100>$. \\

The neutron diffraction studies of Co$_2$MnSi compound were performed at the Canadian Institute for Neutron Scattering in Chalk River on the C2 High Resolution Powder Diffractometer with a wavelength of 1.33{\AA}. The neutron wavelength was calibrated using a Al$_2$O$_3$ sample as a standard. The diffraction data were collected on the 4 gram of the powdered polycrystalline Co$_2$MnSi sample, sealed in a thin-walled vanadium tube under argon atmosphere.\\
The single-crystal diffraction (SCD) was acquired using the \emph{Bruker Smart Apex2 CCD} with Mo K$\alpha_{1}$ radiation on a tiny crystal piece, sizing about 100 $\mu$m. The MAX3D~\cite{Britten2007} software was used to visualize the reciprocal space that confirmed the crystallinity of the material. The crystal structure was solved and refined using the SHELXS and SHELXL~\cite{Sheldrick1997} software packages.\\

The measurement of the magnetization hysteresis loops was performed on a polycrystalline sample with the Quantum Design MPMS SQUID magnetometer in the applied magnetic field up to $\pm4$ Tesla. \\

The electrical resistivity was measured on a single crystal sample of gauge dimensions 6.62 mm$\times$1.6 mm$\times$0.48 mm  with the Keithley 2182A nanovoltmeter and 6221 current source, attached to the Quantum Design PPMS system.  A four-point method of measuring potential drop across the sample was used to determine the sample resistance. The sample was mounted on a platform with a spring-loaded, point-contact potential and current leads. The potential drop was measured in Delta mode as the average of 100 current reversals.  Between 2 K and 298 K, the resistivity measurements were conducted using PPMS platform to control the sample temperature. Between 450 K and 1050 K, the measurements were performed in the resistance furnace inside the quartz tube under argon protective atmosphere. A separately developed ceramic holder with a pressure-loaded, point-contact potential and current leads was used in high temperature resistivity measurements. The temperature was controlled with the thermocouple located in the vicinity of the sample. \\

The dilatometry measurements were carried out on a push-rod dilatometer system on 33 mm long cylindrical single crystal sample with a non-uniform diameter of $\sim$2 mm.\\

The first-principle calculations were carried using the density functional theory (DFT) implemented in the Vienna ab-initio simulation program (VASP) package~\cite{Kresse1993,Kresse1996}, with the generalized gradient approximation (GGA) exchange correlation functional~\cite{Perdew1996}, under the Projector Augmented Wave (PAW)~\cite{Kresse1999} functions. To address the electron localization of 3d states of Co and Mn atoms, GGA+U method~\cite{Dudarev1998} was used to obtain the system equilibrium. Effective Hubbard parameter, U$_{eff}$ of 1.92 eV for Co 3d and 1.62 eV for Mn 3d states was applied which was previously successfully used for Co$_2$MnSi \cite{Kandpal2007}. The formation enthalpy parameters for the optimized structures were computed with the HSE06 hybrid functional that estimate the exchange interaction more efficiently~\cite{Krukau2006}. The Brillouin zone for single elementary cell calculations was sampled using $6\times6\times6$ k-mesh. To study the effect of antisite disorder in Co$_2$MnSi, a $2\times2\times2$ supercell was generated and the k-mesh was adjusted accordingly to maintain the same k-point density.

\section{Results}

\subsection{Single crystal diffraction (SCD)}

Co$_2$MnSi crystalizes with the $L2_{1}$ type crystal structure within the space group $225\,\, Fm\overline{3}m$. The Co, Mn, and Si atoms are situated in the Wyckoff position 8c (3/4, 1/4, 1/4), 4a (0,0,0) and 4b (1/2, 1/2, 1/2), respectively. The refined single crystal structural parameters are summarized in Table~\ref{Table:Co2MnSi_SC_XRD_298K} and ~\ref{Table:Co2MnSi_SC_XRD__atomic298K}. The obtained crystal structure was found to be in good agreement with previously published results~\cite{gladyshevskii1961crystal,Webster1971,Sobczak1976,Buschow1983,Ido1988,Fujii1994} available in the inorganic crystal structure database (ICSD)~\cite{Bergerhoff1987}. Note that, the refined XRD-SCD parameters did not yield the information on the transition metal Co-Mn antisite disorder which is considered as one of the main reasons for the weakening of half-metallicity in Co$_2$MnSi compound~\cite{Ravel2002}. The absence of the very well known disorders can be explained by the similar X-ray scattering power of Co and Mn. Consequently, disorders in the range of $\sim10\%$ that was observed previously by~\citet{Ravel2002} are indistinguishable with XRD.

 \begin{table}[htp]
    \caption{Crystallographic data for Co$_2$MnSi single crystal obtained from refinement of X-ray diffraction data (Mo K$_\alpha$ radiation, 298K)}\label{Table:Co2MnSi_SC_XRD_298K}
     \centering
        \begin{tabular}{l l l l l }
        \hline
        \hline

              &   &  \\
            \hline
                & Refined composition &Mn Co$_{2}$Si \\
                &Space group& $Fm\bar{3}m$ \\
                &Lattice constant $({\AA})$& $5.6585(4)$ \\
                &Volume $({\AA}^{3})$& $181.18(2)$ \\
                &$\rho_{calc} (g/cm^3))$& $7.3642$ \\
                &Z & $4$ \\
                &$2\theta$ range & $12.48 - 88.3$ \\
                &Index ranges& $-7\leq h\leq10$,$-10\leq k\leq11$$-9\leq l\leq11$ \\
                &Reflections collected& $537$ \\
                &Independent reflections& $58$$[R_{int}=0.0168,\,R_{sigma=0.0098}]$ \\
                &Data/restraints/parameters& $58/0/4$ \\
                &Goodness-of-fit on $|F|^2$& $1.185$ \\
                &Largest diff. peak/hole (e/${\AA}^{3}$)& $0.72/-0.59$ \\
                &R indices $[I>=2\sigma(I)]$& R$_1$=0.0175 wR$_2=$0.0487\\

                \hline
                \hline
        \end{tabular}
\end{table}

 \begin{table}[htp]
    \caption{Occupancy and isotropic displacement parameters for Co$_2$MnSi single crystal.}\label{Table:Co2MnSi_SC_XRD__atomic298K}
     \centering
        \begin{tabular}{l l l l l l l l }
        \hline
        \hline
        &Atom& x & y&z& Site&Occupancy&U$_{eq}$ (\AA$^{2}$)\\
        \hline
        &Co&3/4&3/4&1/4&8c&1.00008&0.00406(18)\\
        &Mn&1/2&1/2&0&4a&0.99984&0.00404(19)\\
        &Si&0&0&1/2&4b&0.99984&0.005(3)\\
                \hline
        \end{tabular}
\end{table}

\subsection{Neutron powder diffraction}

In contrast to X-ray, neutron beam offers coherent scattering lengths of 2.53 fm for Co and -3.73 fm for Mn, respectively. As a consequence, antisite disorder between Co and Mn atoms becomes distinguishable. The neutron diffraction data were collected at 4 K, 100 K, and 298 K in an effort to accurately refine the structure, the magnetic moments and observe their variation with decreasing temperature. One of the challenges in performing a precise refinement for Co$_2$MnSi is the lack of paramagnetic state diffraction data that makes it difficult to differentiate between the structural and magnetic reflections. An ideal approach would involve conducting a refinement on data collected above the Curie temperature to obtain structural information which can then be used successively for magnetic structure determination. However, the high Curie temperature of Co$_2$MnSi does not permit the data collection in the paramagnetic state. Furthermore, as discussed earlier, X-ray diffraction can not also successfully capture the Co-Mn disorder accurately which makes refinement more challenging.\\

 \begin{figure}[htp]
\centering
   \includegraphics[scale=0.5]{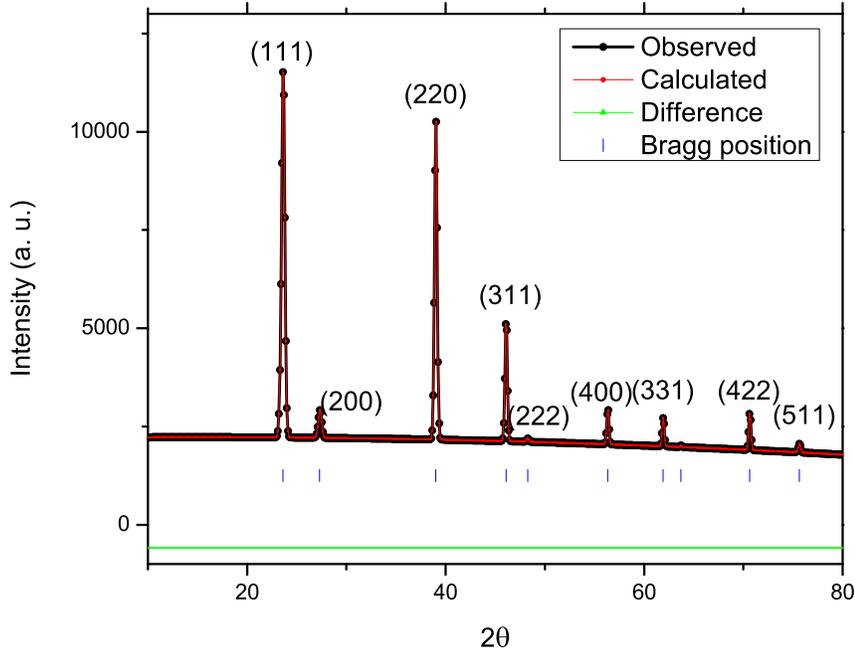}\\
   \caption{Neutron pattern simulation of the magnetic structure with the parameters obtained from the first-principles calculation of defect free Co$_2$MnSi.}\label{fig:magnetic_simulation}
\end{figure}

Simulation of magnetic reflections with the theoretically obtained magnetic moments (Fig.~\ref{fig:magnetic_simulation}) suggest the presence of three dominant reflections (111), (220) and (311) for Co$_2$MnSi. Among these, the intensity of (111) and (310) are also significant to determine the Mn-Si disorder as they fall into the category with h, k, l all being odd~\cite{Webster1971}. As a result, separating these magnetic reflections becomes crucial for a good refinement.\\

\begin{figure*}
\subfigure[\label{sfig:Neutron_298K_high_angle}]{%
  \includegraphics[scale=0.3]{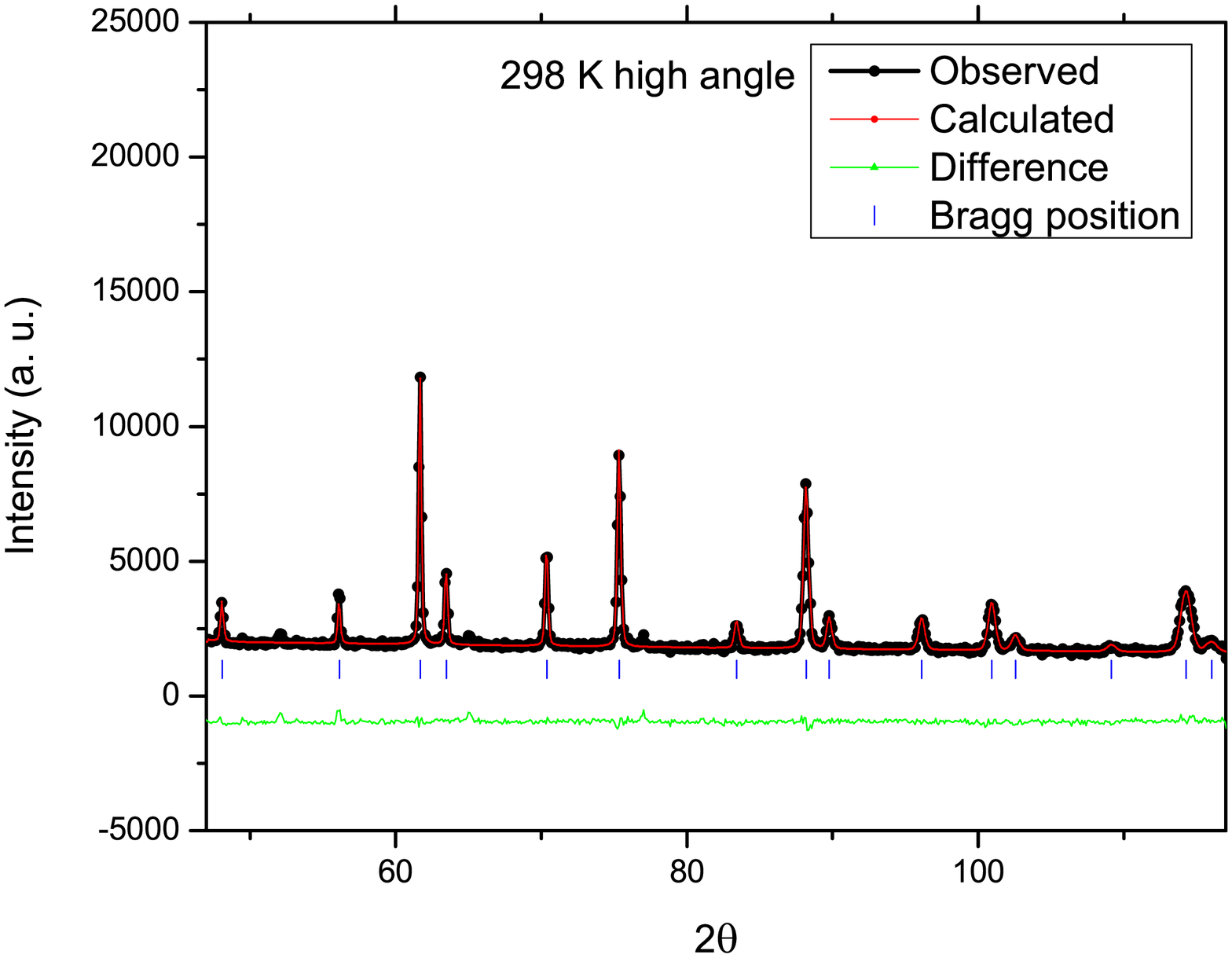}%
}\hfill
\subfigure[\label{sfig:298K_neutron}]{%
  \includegraphics[scale=0.3]{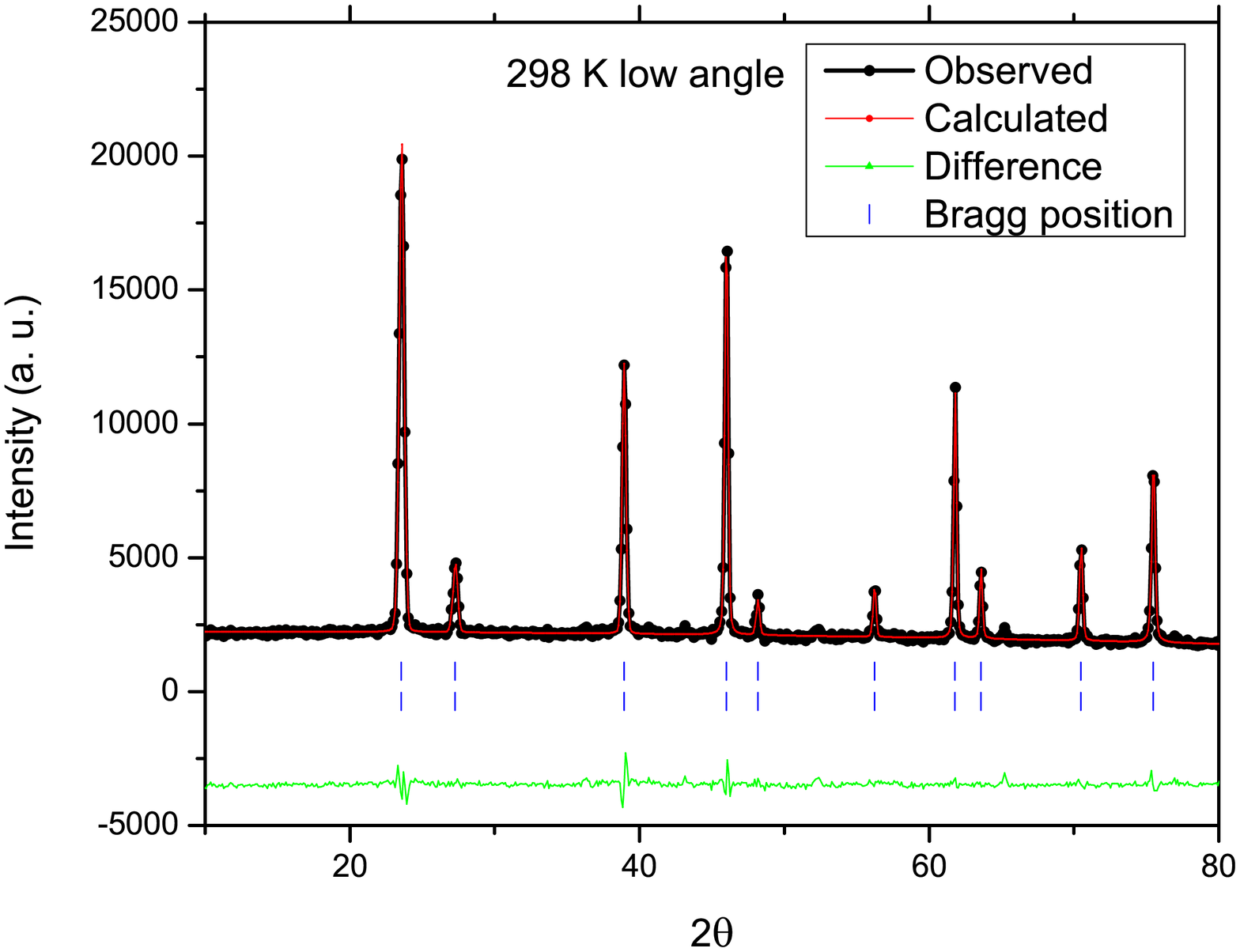}%
}\hfill
\subfigure[\label{sfig:100K_neutron}]{%
  \includegraphics[scale=0.3]{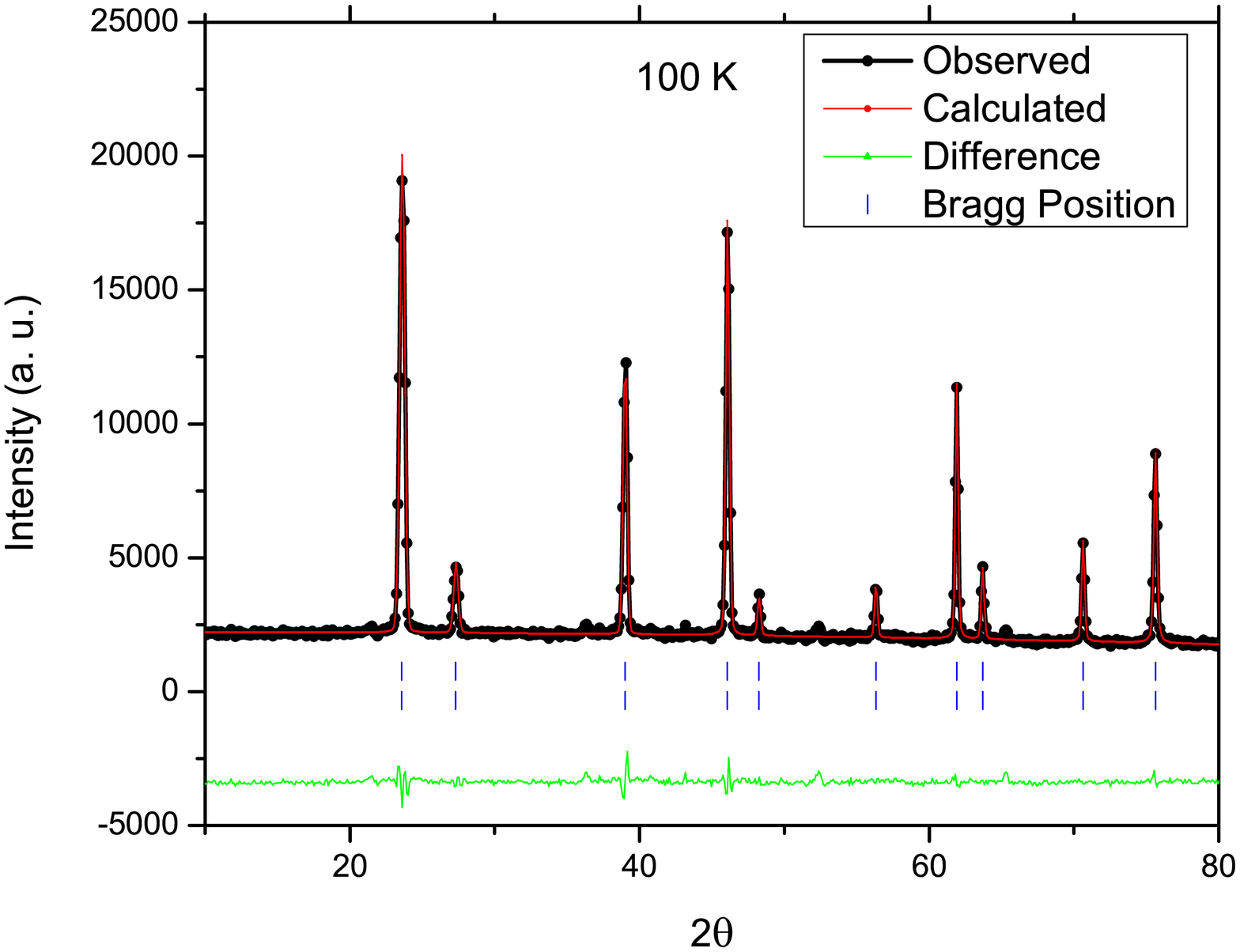}%
}
\subfigure[\label{sfig:4K_neutron}]{%
  \includegraphics[scale=0.3]{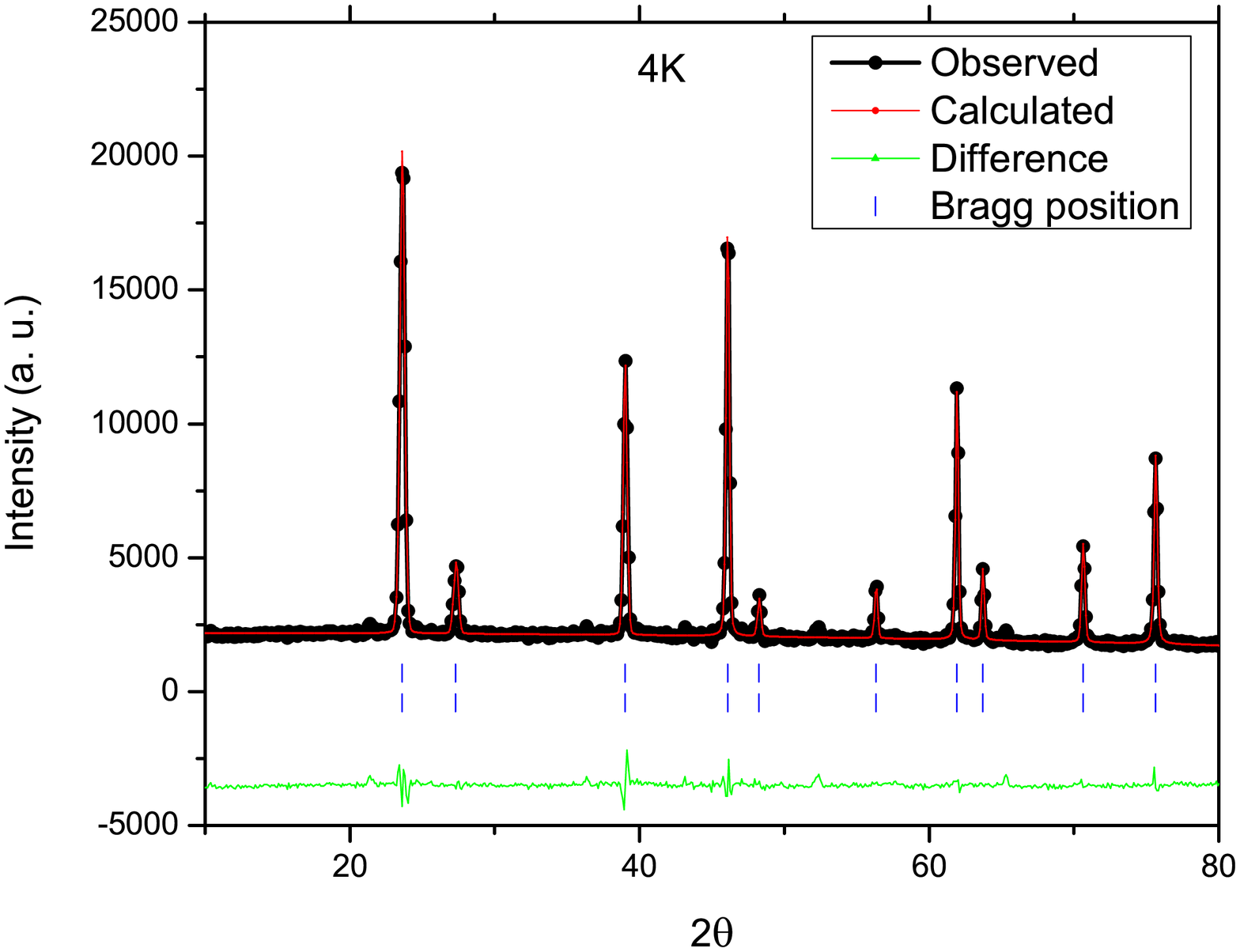}%
}
\caption{Rietveld refinement profile of the neutron powder diffraction data at a) 298 K with high angle structural refinement, and  b) 298 K c) 100 K and d) 4 K structural and magnetic refinement.}
\label{fig:All_neutron}
\end{figure*}

The pattern simulation in Fig. \ref{fig:magnetic_simulation} shows that all the major magnetic peaks lie below the $2\theta$ value of 46.5$^\circ$. If the weak magnetic reflections above 46.5$^\circ$ are considered negligible, it can provide a means to obtain Co-Mn disorder more accurately. In this work, the structural refinement for Co$_2$MnSi was performed at higher angles within the range (47$^\circ$ - 117$^\circ$).  Such a refinement is feasible in the present case because in neutron scattering stronger reflections are observed at higher angles that allow proper identification of elements. The refinement profile is shown in Fig. \ref{fig:All_neutron}a and refined structural parameters are listed in Table~\ref{Table:Co2MnSi_refined}. It can be seen that Mn and Co tend to from antisite disorder among themselves with $\sim$6.5\% (3.25$\times$2) Co sites being occupied by Mn and $\sim$7.6\% Mn sites being replaced by Co. Refinement involving a portion of Mn and Si sitting in each other's sites yielded almost negligible occupancy and was ignored in the subsequent refinement. The obtained occupancy parameters were kept constant and used in the successive magnetic refinements at 298 K, 100 K, and 4K. It should be noted that although the refinement ignored the presence of weaker magnetic reflections at higher angles, their presence in the refinement should be within the limit of experimental errors.\\

\begin{figure}[htp]
\centering
   \includegraphics[scale=0.5]{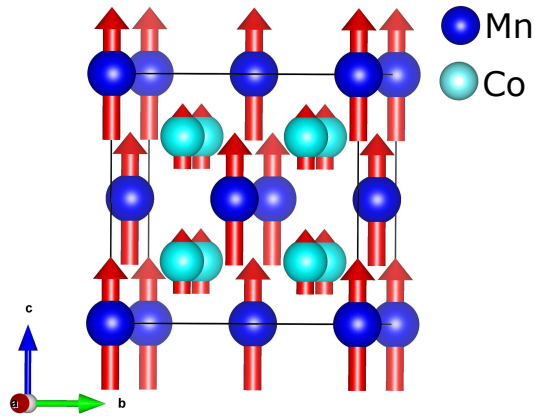}\\
   \caption{Refined magnetic structure of Co$_2$MnSi of 298 K.}\label{fig:Magntic_structure_Co2MnSi}
\end{figure}

The magnetic structure at 298 K was refined with all structural parameters obtained from the higher angle refinement. The initial values of the magnetic moments were taken from our first-principles calculation that will be discussed in an upcoming section. The solution of the magnetic structure was based on the representation analysis approach using SARAh~\cite{Wills2000}. For the ordering, wave vector k = (0 0 0) was used since no extra magnetic reflection was observed. The method yielded only one basis vector, $\Gamma_9$ that corresponds to a ferromagnetic interaction. The final refined structural parameters are listed in Table~\ref{Table:Co2MnSi_refined}, and the refinement profile is shown in Fig. \ref{sfig:298K_neutron}. The refined ferromagnetic structure is displayed in Fig.~\ref{fig:Magntic_structure_Co2MnSi}. Obtained magnetic moment values of 2.432(37) $\mu_B/f.u.$ for $4a$ site (Mn and Disorder Co) and 0.962(30) $\mu_B/f.u.$ for $8c$ (Co and Disorder Mn) are much smaller compared to the previous results at 298 K~\cite{Brown2000,Ravel2002}.  Note that, the neutron magnetic refinement was carried out for the particular magnetic sites rather than individual atoms. This is because the moments associated with disorder of $\sim$6.5\% and $\sim$7.6\% are usually small and can not be solved accurately using neutron diffraction technique. However, a site specific magnetic refinement captures the contribution from both the small antisite disorder and the parent occupant.\\

The refinement of the 100 K and 4 K neutron diffraction data were also performed with the occupancy parameters obtained from 298 K high angle structural refinement. No additional reflections were observed at low temperature revealing lack of any structural or magnetic transition below 100 K. The refinement profile is shown in Fig. \ref{sfig:100K_neutron} (100 K) and \ref{sfig:4K_neutron} (4 K), respectively. The relevant structural and magnetic parameter are listed in Table~\ref{Table:Co2MnSi_refined}. The lattice constant of Co$_2$MnSi is less sensitive to the temperature, which is apparent when comparing their values at different temperatures. The magnetic moments are increased as the temperature is lowered. Nevertheless, the refined values at 4 K are much lower compared to the results of ~\citet{Webster1971} at the same temperature.

\begin{table}[htp]
    \caption{Refined structural parameters for Co$_2$MnSi}\label{Table:Co2MnSi_refined}
     \centering
        \begin{tabular}{c c c c c c c}
        \cline{1-6}\\
        \cline{1-6}
        &Temperature& 298 K-high angle&298 K &100 K&4 K\\
     \cline{1-6}
        &Spacegroup&\multicolumn{4}{c}{Fm$\bar{3}$m}\\
        &Lattice constant(${\AA}$)&5.6406(4)&5.6406&5.6301(7)&5.6301(2)\\
        &Magnetic phase&-& \multicolumn{3}{c}{Ferromagnetic}\\
        \cline{1-6}
        &\multicolumn{4}{c} {Co and disorder Mn; and  8c (3/4, 3/4, 1/4)}\\
        &Occupancy (Co)& \multicolumn{4}{c}{0.9675 (28)}\\
        &Occupancy (Mn)& \multicolumn{4}{c}{0.0325 (28)}\\
        &B(${\AA}^2$)&0.607(2)&0.607&0.4318(13)&0.3565(5)\\
        &M($\mu_B/f.u.$)&-&0.962(30)&1.008(16)&0.906(18)\\
        \cline{1-6}
        &\multicolumn{4}{c} {Mn and Disorder Co; 4a (1/2, 1/2, 0)}\\
        &Occupancy (Mn)& \multicolumn{4}{c}{0.924 (43)}\\
        &Occupancy (Co)& \multicolumn{4}{c}{0.076 (43)}\\
        &B(${\AA}^2$)&0.6877(1)&0.6877&0.4759(5)&0.5561(15)\\
        &M($\mu_B/f.u$)&-&2.432(37)&2.456(32)&2.62(20)\\
        \cline{1-6}
        &&\multicolumn{4}{c} {Si; 4b (0, 0, 1/2)}\\
        &Occupancy (Si)& \multicolumn{4}{c}{1}\\
        &B(${\AA}^2$)&1.0177(14)&1.0177&1.1147(27)&1.1114(16)\\
        \cline{1-6}
        &$\chi^2$&2.92&5.64&4.44&5.57\\
        &$R_{wp}$&14.5&13.6&13&13.1\\
        &$R_{F}$&4.38&4.58&6.16&6.23\\
        &$R_{mag}$&-&3.57&2.37&3.64\\
        \cline{1-6}
        \end{tabular}
\end{table}


\subsection{Magnetization behaviour}

\begin{figure}[htp]
\centering
   \includegraphics[scale=0.5]{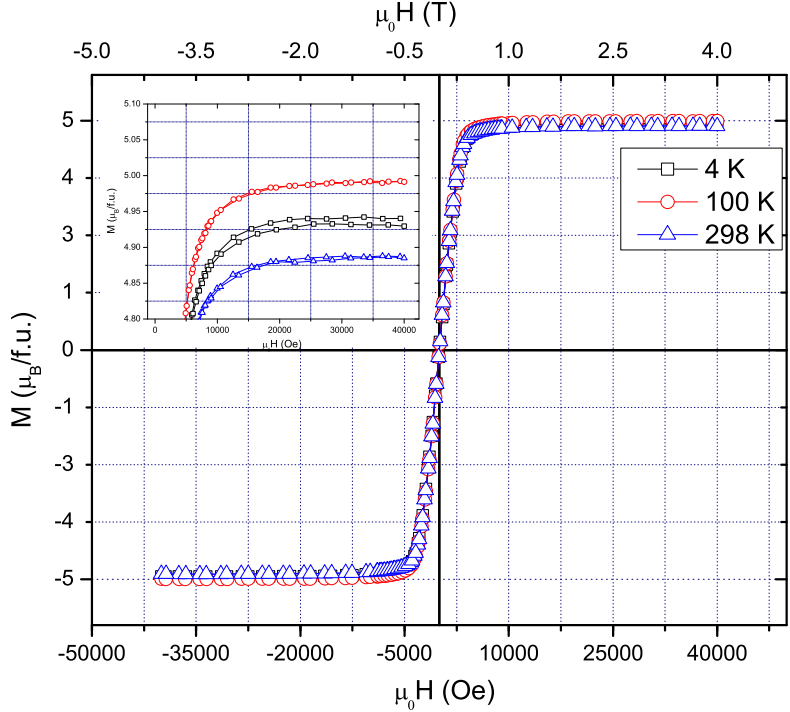}\\
   \caption{Magnetic hysteresis loops at 4 K, 100 K and 298 K, measured with the applied field up to 40 kOe (4 Tesla). Inset shows the saturation magnetization at the corresponding temperatures.}\label{fig:Co2MnSi_hysteresis}
\end{figure}

The hysteresis behaviour of the Co$_2$MnSi compound at 4 K, 100 K and 298 K is shown in Fig.~\ref{fig:Co2MnSi_hysteresis}. The compound exhibits almost identical saturation magnetization of $(M_{S})$ 4.94 (1), 4.99 (1) and 4.89 (1) $\mu_{B}/f.u.$ at 4 K, 100 K, and 298 K, respectively. Previous experimental measurements also reported comparable results of 4.96 $\mu_{B}/f.u.$~\cite{Brown2000} and 5.15 $\mu_{B}/f.u.$~\cite{Ravel2002}. The coercive field was found to be $\sim$20 (5) Oe for all the three temperatures of the study.\\

\subsection{Dilatometric measurements}
\begin{figure}[htp]
\centering
   \includegraphics[scale=0.5]{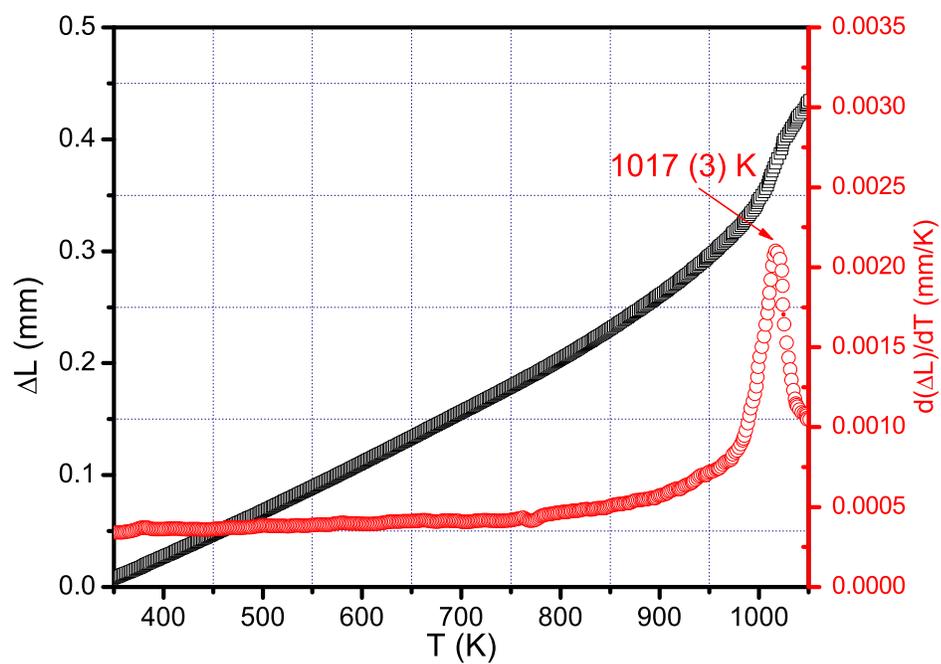}\\
   \caption{$\Delta$L and $\frac{d(\Delta L)}{dT}$ as a function temperature measured from 350 K to 1150 K. $\Delta$L shows a linear response with temperature until near the magnetic phase transition point where the response changes. The Curie temperature was identified to be 1017 (3) K from the sharp peak in the $\frac{d(\Delta L)}{dT}$ vs T plot.}
   \label{fig:Dilatometry}
\end{figure}

The dilatometric measurements were performed on a single crystal sample to determine the Curie temperature. The technique has been used previously to determine the ferromagnetic to paramagnetic transition temperature of several compounds~\cite{Liu2004,Mohapatra2007,Mohapatra2007a,Verma2010}. The change of length, $\Delta$L and its derivative with respect to temperature, $\frac{d(\Delta L)}{dT}$, as a function of temperature for Co$_2$MnSi is shown in the Fig.~\ref{fig:Dilatometry}. It can be seen that the length change starts to deviate from a linear responsive behaviour at higher temperatures where the ferromagnetic to paramagnetic phase transition takes place. From the sharp peak in the $\frac{d(\Delta L)}{dT}$, the Curie temperature was identified to be 1017 (3) K which compares well with the previously published $T_{C}$ of 985 K~\cite{Webster1971,Brown2000}

\subsection{Transport properties}

The total electrical resistivity of Co$_2$MnSi compound consists of three main contributions, the residual resistivity ($\rho_{R}$), magnonic ($\rho_{M}$) and phononic parts ($\rho_{P}$), as per Eq.\ref{Eqn:Total_resisitivity}~\cite{Bombor2013}.

\begin{equation}\label{Eqn:Total_resisitivity}
\rho=\rho_{R}+\rho_{M}(T)+\rho_{P}(T).
\end{equation}

The residual resistivity, $\rho_{R}$, is determined by the imperfections in the crystal. The magnetic resistivity, $\rho_{M}$, takes into account the exponential suppression of the quadratic temperature dependence due to magnon scattering  ~\cite{Campbell1982,Goodings1963,barry1998evidence,Bombor2013} and is given by (Eq.\ref{Eqn:Magnonic_resisitivity}):

\begin{equation}\label{Eqn:Magnonic_resisitivity}
\rho_{M}(T)=AT^{2}e^{\frac{-\Delta}{T}}.
\end{equation}
here, the constant $A$ indicates the strength of magnon scattering and $\Delta$ is the minority spin gap at the Fermi level, $E_{F}$.\\

The phononic part of the electrical resistivity, $\rho_{P}$, is analyzed by the Bloch-Gr$\ddot{u}$neisen formula (Eq.\ref{Eqn:Bloch-Gruneisen_resisitivity}).

\begin{equation}\label{Eqn:Bloch-Gruneisen_resisitivity}
\rho_{P}(T)=C({\frac{T}{\theta_{D}}})^{n}\int^{\frac{\theta_{D}}{T}}_{0}\frac{x^{n}}{(e^{x}-1)(1-e^{-x})}
\end{equation}
where A is constant and $\theta_{D}$ is Debye temperature. A $\theta_{D}$ value of 520 K was reported for Co$_{2}$MnSi by~\citet{Ito2017}. The parameter $n$ can be either 3 or 5 that identifies mechanism of phononic scattering. A fitting with $n=3$ indicates an interband electron scattering from conduction band, while $n=5$ corresponds to a intraband scattering.\\

\begin{figure*}
\subfigure[\label{sfig:Resistivity-Low-temperature}]{%
  \includegraphics[scale=0.3]{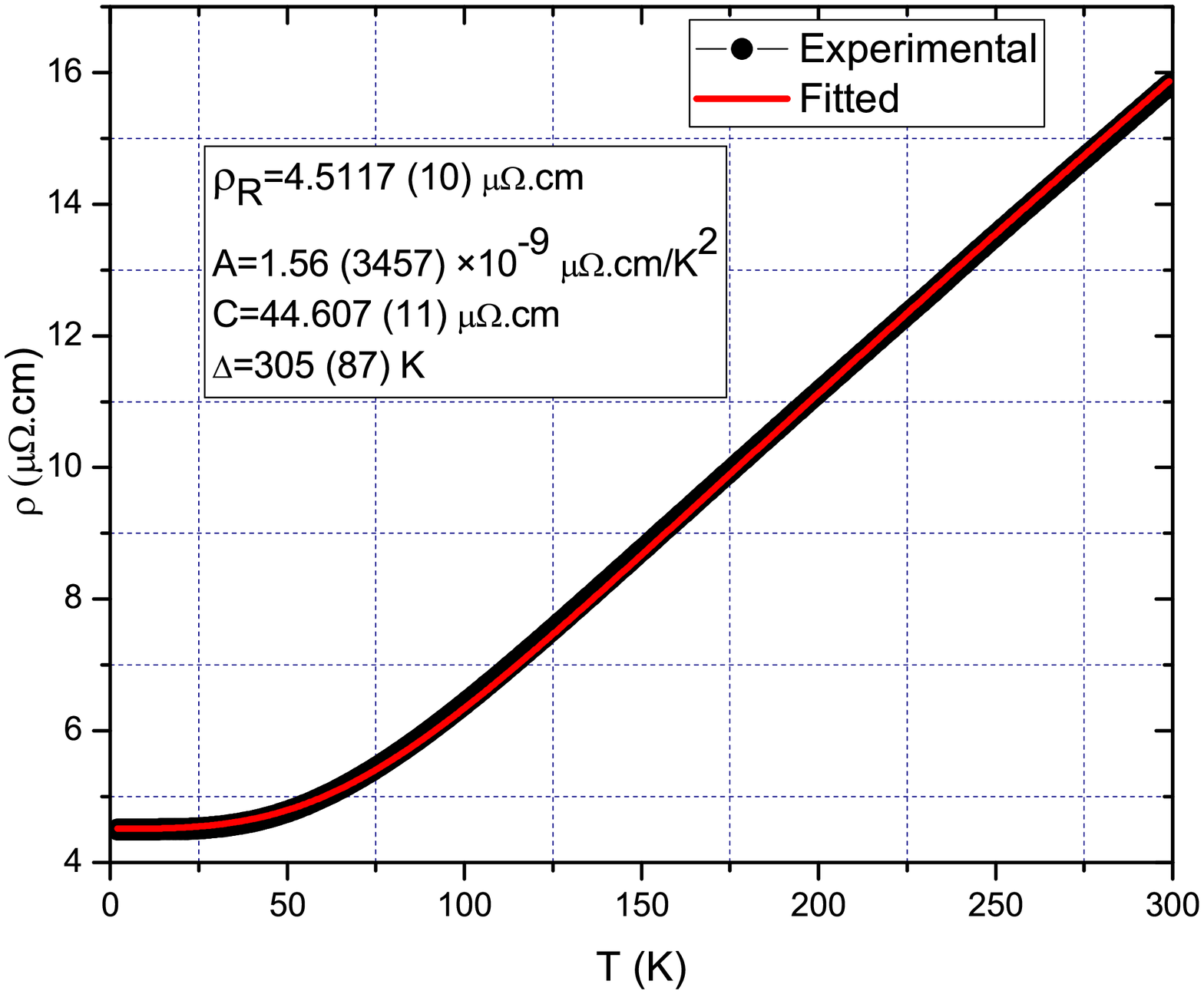}%
}
\subfigure[\label{sfig:Resistivity-High-temperatrure_derivative}]{%
  \includegraphics[scale=0.3]{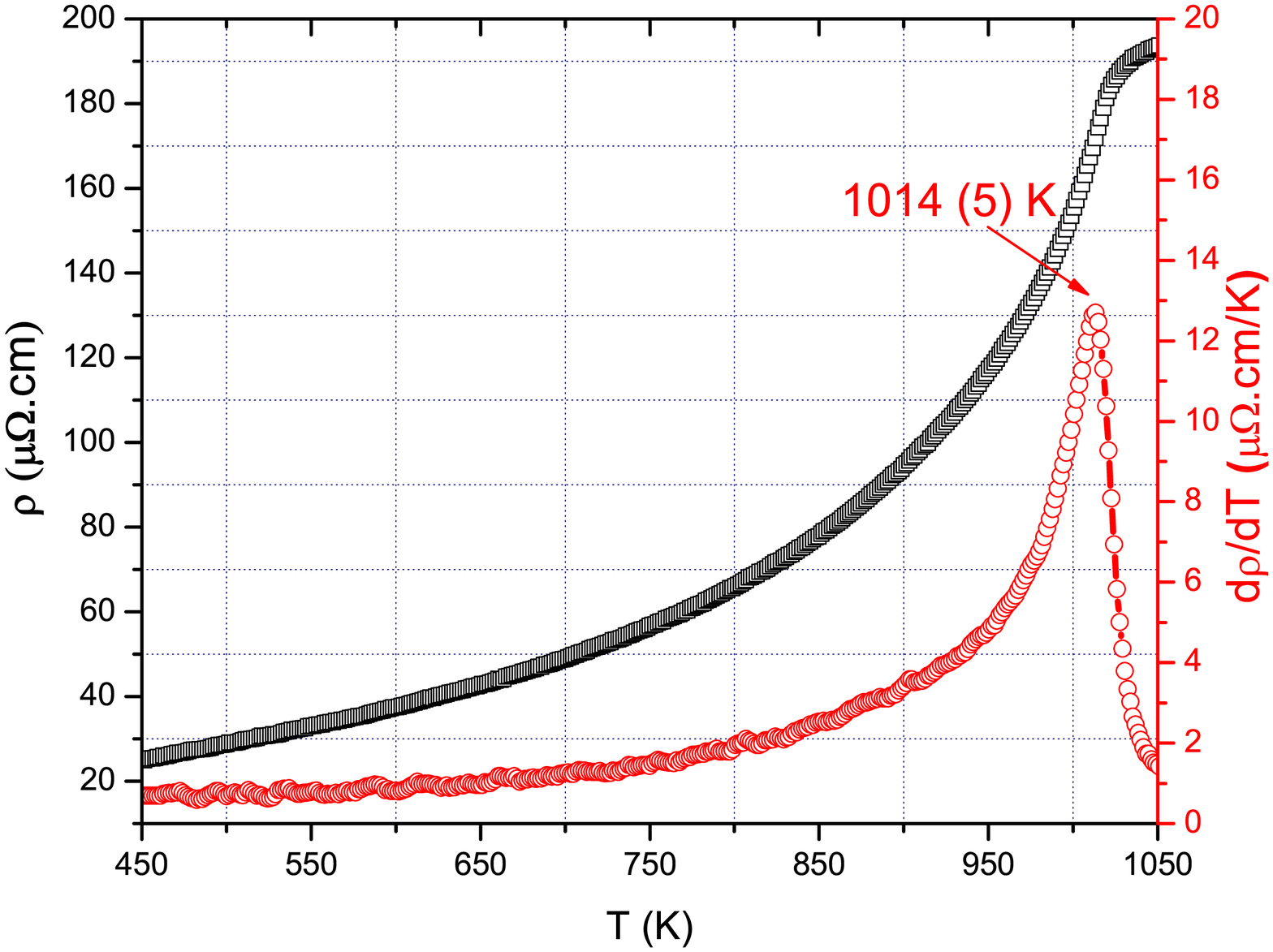}%
}\hfill
\caption{a) Low temperature (2 K to 300 K) electrical resistivity of Co$_2$MnSi single crystal fitted using Eq.~\ref{Eqn:Total_resisitivity}. High temperature electrical resistivity, $\rho$ and $\frac{d\rho}{dT}$, as a function of temperature measured from 450 K to 1150 K. The Curie temperature was determined to be 1014 (5) K from the sharp change in the temperature dependence of $\frac{d\rho}{dT}$.}
\label{fig:Co2MnSi_resisitivity}
\end{figure*}

For Co$_{2}$MnSi, the best fitting was obtained using $n=3$ in the Bloch-Gr$\ddot{u}$neisen expression (Eq.~\ref{Eqn:Bloch-Gruneisen_resisitivity}), as it is shown in Fig.~\ref{sfig:Resistivity-Low-temperature}.  The residual resistivity, $\rho_{R}$ was determined to be 4.5517 (10) $\mu\Omega$.cm, which is in good agreement with previously published values of 2$\mu\Omega$cm~\cite{Raphael2002} and 7$\mu\Omega$cm~\cite{Ritchie2003}. It can be seen from the fitting parameters that electrical resistivity from 2 K to 300 K is completely dominated by the interband phonon scattering. The significantly small value 1.56 (2457)$\times$ 10$^{-9}$ $\mu\Omega$.cm/K$^{2}$ of parameter $A$ with higher errors indicates the magnon scattering to be almost negligible upto 300 K which also makes the obtained $\Delta$ of 305 (87) K unreliable. A temperature independent behavior was observed below 30 K, indicating the system to be a good metal at low temperatures. The residual resistivity ratio (RRR) at 300 K, i.e., the ratio of the resistance of the sample at 300 K over the resistance at 2 K, was found to be 3.5.\\

The Curie temperature was determined from the high temperature resistivity measurement with an apparent sudden change in the electrical resistivity at higher temperatures (Fig.~\ref{sfig:Resistivity-High-temperatrure_derivative}). From the peak of the $\frac{d(\rho)}{dT}$ vs. T plot, the Curie temperature was identified to be 1014 K which corresponds well to the dilatometry measurements. The RRR value for Co$_2$MnSi reaches $\sim$ 50 near the transition temperature.\\

\subsection{Magnetoresistance}

\begin{figure*}
\subfigure[\label{sfig:Resistivity_temp_field}]{%
  \includegraphics[scale=0.3]{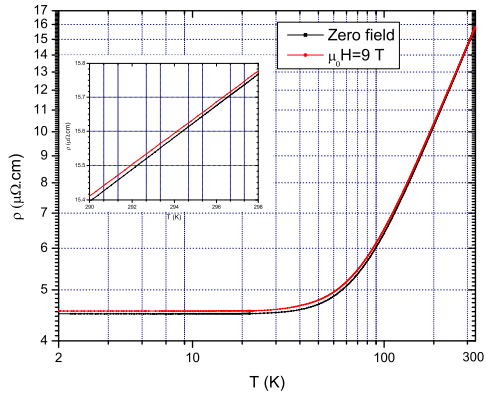}%
}\hfill
\subfigure[\label{sfig:Temp_MR_9T}]{%
  \includegraphics[scale=0.3]{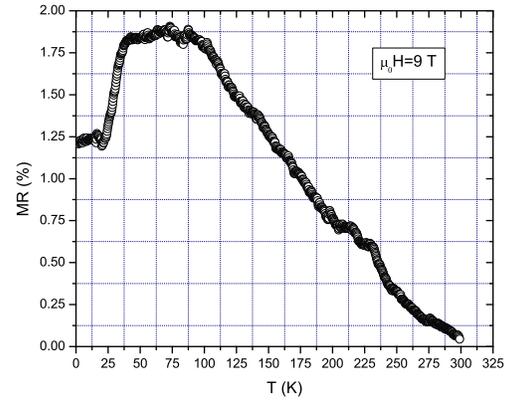}%
}\hfill
\subfigure[\label{sfig:Field_MR_2K}]{%
  \includegraphics[scale=0.3]{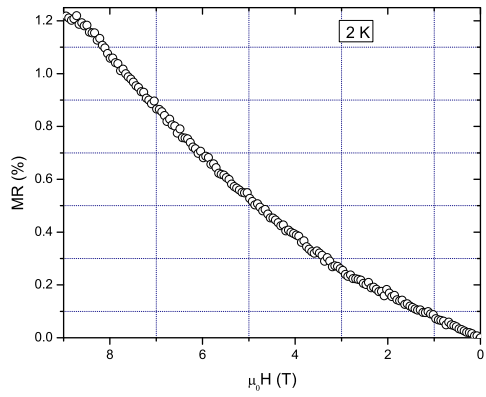}%
}
\caption{a) Temperature dependence of electrical resistivity in a zero (black) and 9 Tesla magnetic field (red), showing a very small increase in electrical resistivity under the external field. Inset shows the comparison of the data near room temperature. b) Change in magnetoresistance, MR (\%), as a function of temperature. c) Magnetoresistance at 2 K as a function of magnetic field measured from the saturated state at 9 Tesla down to 0 Tesla.}
\label{fig:All_magnetoresistance}
\end{figure*}

The magnetotransport of Co$_{2}$MnSi compound was measured between 2 K and 302 K, in the constant magnetic field of 9 Tesla applied perpendicular to the direction of current flow (transverse magnetoresistance). In comparison with the zero field data, Co$_{2}$MnSi shows a very small and positive increase in magnetoresistance up to room temperature (Fig.~\ref{sfig:Resistivity_temp_field}). The magnetoresistance, MR (\%)$(=\frac{\rho(9\,T)-\rho(0\,T)}{\rho(0\,T)}\times 100)$, as a function of temperature ((Fig.~\ref{sfig:Temp_MR_9T})) shows a maximum value of 1.88(1)\% at 68 K, which decreases to 0.06(1)\% at 300 K. Fig.~\ref{sfig:Field_MR_2K} shows the field dependence of the electrical resistivity at 2 K. The resistance of Co$_{2}$MnSi increases roughly parabolically with the applied field, with the value of MR $\sim 1.22(1)\%$ at the highest field of 9 Tesla available in our system (Fig.~\ref{sfig:Field_MR_2K}).\\

\subsection{First-principle calculations}

Calculations were carried out on a pure Co$_2$MnSi, as well as the structures with different concentrations of disorder that are relevant to the experimental observations. Before computation of the energy, a complete first-principles structural relaxation was performed. Here, we consider various Co-Mn antisite disorders and discuss their propensity to form in terms of the energy difference with the ideal Co$_2$MnSi. Note that these calculations do not include any disorder involving the Si (4b) site as they have not been observed in any previous experimental investigations~\cite{Webster1971,Brown2000,Ravel2002}, nor in the present work. Four different types of defects were studied with all the calculations done with a 128 atoms unit cell. The first one considers a Mn only antisite disorder created by replacing one Co atom to obtain a (Co$_{0.984}$Mn$_{0.016}$)$_2$MnSi stoichiometry. The second one corresponds to a Co only antisite with a composition of (Co$_2$(Mn$_{0.969}$Co$_{0.031}$)Si. The third and fourth type of defects involves disorder between both Co-Mn at concentrations that are relevant to the experimental results. It should be noted that the neutron refinement in this study and other experimental investigations indicated slightly different disorder affinity for the Mn and Co sites. In the first-principle calculations, however, we adjusted the identical disorder occupancy between Co and Mn to preserve accurate stoichiometry. The computed energies are listed in Table~\ref{Table:Defect_Formation}. The computed defect formation energies included the addition and subtraction of the excess pure Co and Mn from the reservoirs.\\

 \begin{table}[htp]
    \caption{Defect formation energies for the disordered configurations in the 128 atoms supercell Co$_2$MnSi. The energy for the ideal structure without any defect is taken as 0.}\label{Table:Defect_Formation}
     \centering
        \begin{tabular}{l l l l l l l}
        \hline
        \hline

              &Type of disorder&Composition &   Required energy (eV/f.u.)&    \\
            \hline
                &Defect free&Co$_2$MnSi & 0\\
                &Only Mn antisite&(Co$_{0.984}$Mn$_{0.016}$)$_2$MnSi & 0.041\\
                &Only Co antisite&(Co$_2$(Mn$_{0.969}$Co$_{0.031}$)Si & 0.30 \\
                &Co-Mn swap&(Co$_{0.969}$Mn$_{0.031}$)$_2$(Mn$_{0.938}$Co$_{0.062}$)Si & 0.015  \\
                &Co-Mn swap&(Co$_{0.953}$Mn$_{0.047}$)$_2$(Mn$_{0.906}$Co$_{0.094}$)Si& 0.073  \\
                \hline
                \hline
        \end{tabular}
\end{table}

From Table~\ref{Table:Defect_Formation}, it can be seen that formation of a single Mn antisite defect (1.6\%) in Co$_2$MnSi requires external energy of 0.041 eV/f.u while the single Co counterpart (3.1\%) requires 0.3 eV/f.u. The calculation is consistent with the work of \citet{Picozzi2004}, which showed the Co only antisite disorder is energetically less favourable. Formation of such single antisite defects requires, however, an excess supply of Mn/Co from the reservoir and expelling of an equivalent concentration Co/Mn from the ideal structure that break the ideal 2:1:1 stoichiometry. In comparison, as can be seen in Table~\ref{Table:Defect_Formation}, equivalent Mn-Co antisite disorders with higher defect concentrations requires much less energies. In our calculations, the formation of 6.2\% and 9.4\% antisite Co-Mn disorder was requires 0.015 eV/f.u. and 0.073 eV/f.u, respectively. These correspond to thermal energies at 173 K and 846 K, which are usually encountered during the experiments. Furthermore, these equivalent Co-Mn swaps maintains the ideal 2:1:1 stoichiometry and thus are more favourable to form, in contrast to the Mn or Co only antisite disorder. Note that our computed energies are in good agreement with the reported energies by~\citet{Pradines2017} with similar defect concentrations.\\


\begin{table}[htp]
\begin{center}
     \caption{Average magnetic moment from the first-principle calculations. Column "A" stands for the (Co$_{0.969}$Mn$_{0.031}$)$_2$(Mn$_{0.938}$Co$_{0.062}$)Si composition. Column "B" describes (Co$_{0.953}$Mn$_{0.047}$)$_2$(Mn$_{0.906}$Co$_{0.094}$)Si compound.}
       \label{Table:Co2MnSi_dft_magnetic moment}
    \centering
        \begin{tabular}{c c c c c c c c c c}
        \hline
        \hline
        &Atom&Site& \multicolumn{3}{c}{Magnetic Moment ($\mu_{B}/f.u.$)}\\
        \hline
        &&&Co$_2$MnSi&A& B\\
        \hline
        &Co&8c&0.974&1.138&1.114\\
        &Mn disorder&8c&-&-0.111&-0.164\\
        &Mn&4a&3.49&2.666&2.558\\
        &Co disorder&4a&-&0.139&0.18\\
        &Si&4b&-0.102&-0.072&-0.069\\
        &Interstitial region&&-0.298&-0.256&-0.257\\
        &Net moment&&5.04&4.533&4.311\\
                \hline
        \end{tabular}
        \end{center}
\end{table}

In the next step, we considered the effect of defect concentrations on the magnetic properties of Co$_2$MnSi. Here, the ideal Co$_2$MnSi was compared with the structure containing 6.2\% and 9.4\% Co-Mn antisite disorder, discussed above. The magnetic moments are listed in Table~\ref{Table:Co2MnSi_dft_magnetic moment}. From Table~\ref{Table:Co2MnSi_dft_magnetic moment}, it is evident that the net magnetic moment is significantly reduced with the introduction of the antisite disorder. This suppression is caused by an antiferromagnetic interactions of the Mn antisite disorder with the parent Mn atoms, which was also reported in previous theoretical examinations~\cite{Picozzi2004,Pradines2017}. The emergence of these interactions can be attributed to the reduction of Mn-Mn interatomic distances due to the disordered Mn replacing Co atoms in the 8c site~\cite{Pradines2017}. In the previous studies by~\citet{Pradines2017}, it was found that introduction of 6.2\% disorder reduces the average magnetic moment of Mn atoms in the 4a to 2.667 $\mu_{B}/f.u.$ with some Mn atoms coupling antiferromagnetically with the disorder Mn in 8c. As the concentration of disorder increases, more Mn atoms occupy the 8c sites. Consequently, it is expected that antiferromagnetic interaction will be stronger and the reduction of both Mn and hence the total magnetic moment will be higher, which is reflected in the values for 9.4\% Co-Mn swap.

\section{Discussion}

\subsection{Occupancy of Co and Mn antisite disorder}

The neutron diffraction data were collected on a polycrystalline sample that was water quenched from 1073 K (800$^\circ$C) to freeze the state of the system at that temperature. The refinement of the data reveal that the occupancies of the Mn and Co antisite disorder in the state at 1073 K are $\sim$6.5\% and $\sim$7.6\%, respectively (Table~\ref{Table:Co2MnSi_refined}). In contrast, the first-principles calculation showed that the formation of 6.2\% and 9.4\% equivalent Co-Mn disorder requires temperature of 173 K and 846 K, respectively. The disagreement between the theoretical and experimental observations can be attributed to several experimental and theoretical error factors. Some imbalance can originate from the smaller lattice constant obtained theoretically, which in turn produces slightly lower energy compared to the energy that would be obtained with experimentally synthesized structures~\cite{Pradines2017}. Additionally, an error may also be introduced by the experimentally refined antisite disorder with dissimilar amounts of Co and Mn. In contrast, our first-principles calculations suggest the formation of disorder with identical Co and Mn concentrations is more favorable. Other experimental error factors that are difficult to control are related to: (i) the contribution from weak magnetic reflections at higher 2$\theta$ angle data, (ii) evaporation of Mn during arc melting, which is a  common problem in synthesis of Mn based compounds, (iii) purity of the elements used to produce Co$_2$MnSi, (iv) the exact quenching temperature being less than 800$^\circ$C due to the evacuated quartz tube. All these factors can contribute to the observed slight difference in the theoretical and experimental disorder occupancy. Nonetheless, disorder of 9.4\% at 846 K calculated from the first principles and the disorder of $\sim$6.5-7.6\% obtained from the neutron refinement of sample annealed at 1073 K can be regarded to be in fair agreement with each other.

\subsection{Influence of disorder on magnetic properties}

The first principle calculations predict a total magnetic moment of 4.53 $\mu_{B}/f.u.$ for 6.2\% disorder and of 4.31 $\mu_{B}/f.u.$ for 9.4\% disorder, compared to a moment of 5.04 $\mu_{B}/f.u.$ for defect-free structure. Such suppression of the magnetic moment with the increasing degree of disorder arises from the antiferromagnetic interactions between Mn atoms that decrease their moment. The influence of the antiferromagnetic Mn atoms is also reflected in the neutron diffraction refinement with a total moment of $\sim$4.356 at 298 K, 4.472 $\mu_{B}/f.u.$ at 100 K and 4.432 at 4 K $\mu_{B}/f.u.$. The magnetic moments for the $8c$ site that contains Co and disordered Mn were found to be 0.906(18) $\mu_{B}/f.u.$ at 4 K. In contrast, the first-principles calculation that corresponds to a 0 K condition yielded moments of 1.027~$\mu_{B}/f.u.$ (1.138-0.111) and 0.95~$\mu_{B}/f.u.$ (1.114-0.164) for the 6.2\% and 9.4\% disordered $8c$ site, respectively. These results are considered to be in excellent agreement with the experimental observation. The refined moments for the $4a$ sites at 4 K from neutron diffraction were  found to be 2.62(20)~$\mu_{B}/f.u.$, which is also in good agreement with the theoretically obtained 2.805~$\mu_{B}/f.u.$ (2.666+0.139) for 6.2\% disorder and 2.738~$\mu_{B}/f.u.$ (2.555+0.18) for 9.4\% disorder.\\

The measurements of the magnetization hysteresis show a saturation magnetization of 4.94 (1), 4.99 (1) and 4.89 (1) $\mu_{B}/f.u.$ at 298 K, 100 K and 4 K, respectively. These figures are very close to the theoretically predicted value of the total moment for defect-free compound, 5.04 $\mu_{B}/f.u.$, where all atomic moments are ferromagnetically aligned in the applied field of 4 Tesla. Such results are not surprising as they indicate that the antiferromagnetic interactions between the Mn atoms are relatively weak in nature.

\subsection{Magnetotransport}

Due to the negligible magnon scattering, an accurate determination of the energy gap parameter $\Delta$, from the fitting the temperature dependent resistivity data in Fig.~\ref{sfig:Resistivity-Low-temperature}, was not possible. Nevertheless, the presence of a near zero positive magnetoresistance at 298 K (Fig.~\ref{sfig:Resistivity_temp_field}), indicates the $\Delta$ is close to room temperature. An exponential decrease of the electron magnon scattering is expected for a half-metallic system with one spin state~\cite{Bombor2013}. Very small value of $A$ constant obtained from the resistivity fitting in Fig.~\ref{sfig:Resistivity-Low-temperature} indicates that electron magnon scattering is becoming weaker as the temperature decreases and vanishes above 298 K. The presence of an exponential factor is indeed visible in the high-temperature resistivity data in Fig.~\ref{sfig:Resistivity-High-temperatrure_derivative}. The positive magnetoresistance in the system is dominated by the conventional cyclotron effect and the scattering of the electrons on phonons and impurities~\cite{TremoletdeLacheisserie2002}. Note that, the absence of magnetoresistance in the study of ~\citet{Ritchie2003} is likely due to the longitudinal magnetoresistance being much weaker in the magnitude than the transverse magnetoresistance measured here.

\section{Conclusions}

We performed an experimental and theoretical study of the structural disorder and its influence on the magnetic properties in the half-metallic Co$_2$MnSi compound. The results suggest that the system consists of equivalent concentrations of disorders involving Co and Mn sites. From the neutron diffraction studies, the Co and Mn disorder of $\sim$6.5\% and $\sim$7.6\% was obtained, which was corroborated by the theoretical study. The results indicate that the antisite disorder is unavoidable in Co$_2$MnSi since the formation of such disorder is energetically favorable in ambient conditions. Antiferromagnetic interactions due to disordered Mn atoms were detected by the first-principles calculations and were correlated with a reduction of the magnetic moments obtained from the refinement of neutron diffraction data at 298 K, 100 K, and 4 K. Co$_2$MnSi compound shows positive magnetoresistance, which decreases as temperature increases to room temperature. The system exhibits a ferromagnetic to paramagnetic transition at $\sim$1014 K, which was determined by the high-temperature dilatometry and electrical resistivity measurements.

\section*{\center{Acknowledgements}}
The authors thank Dr. O. Rubel for the help with first-principles calculations. Financial support of Natural Sciences and Engineering Research Council of Canada under the NSERC Discovery Grant: "Artificially Structured Multiferroic Composites based on the Heusler alloys" is gratefully acknowledged.

\end{document}